\documentclass[12pt]{article}
\usepackage[cp1251]{inputenc}
\usepackage[russian]{babel}
\usepackage[T2A]{fontenc}
\begin{document}
\begin{center}{\bf Geometrization of quantum mechanics}\end{center}
\begin{center}{O.A. Olkhov}\end{center}
\begin{center}{Institute of Chemical Physics, Russian Academy of Sciences,Moscow}\end{center}
The hypothesis is suggested that the equation for the Dirac free
wave field is, in fact, a group-theoretical relation describing
propagation of specific microscopic deviations of space geometry
from the euclidean one (closed topological manifolds). The Dirac
equation for a hydrogen atom can also be interpreted as a relation
that accounts for the symmetry properties of a piece of curved
space. Within the framework of this concept, atoms have no any
pointlike particles (electrons) inside, and the gauge invariance of
electromagnetic field proves to be the natural consequence of the
basic principles of the proposed geometrical approach.

{\bf Introduction}

Geometrization of the gravitational field within the general
relativity theory gave rise to a hope that all physical phenomena
can be reduced to the geometrical notions of curved space-time. Main
efforts were firstly directed toward finding the geometrical
interpretation of the electromagnetic field as a manifestation of
the space-time curvature. These numerous attempts were not
successful, and we note in this presentation only the well known
work of Weyl, who showed that electromagnetic potentials can be
interpreted as connectivities in some special noneuclidean space
(Weyl space)[1]. Then (as in general relativity), Weyl and Fock
attempted to identify the geometry of this space (curvature and so
on) with the geometry of a real space-time distorted due to the
presence of electromagnetic field [1,2]. This hypothesis turned out
to be contradictory to observable proprieties of the real physical
space-time, and the Weyl's results were afterwards considered as the
methodical ones. However, we will show below that the Weyl theory
acquires new important physical meaning within the suggested
topological approach. There were also investigations in succeeding
years in which the potentials of electromagnetic field and of other
gauge fields were interpreted as connectivities in special
"internal" spaces that can be considered as some kind of fibre
spaces, but these spaces have nothing to do with the curvature of
the real space time [3,4].

Attempts to geometrize "matter" were not so numerous as the ones for
electromagnetic and other gauge fields. By the matter we mean
classical particles, quantum particles of nonrelativistic quantum
mechanics, or quantized or nonquantized Fermi fields of the quantum
field theory. These attempts were also not successful, and we note
here only Wheeler's investigations, because he used (as we do)
topological approach for his geometrization of the classical
electric charges [5]. As for the geometrization of the quantum
mechanics, we have to keep in mind the exceptional accuracy of the
modern quantum formalism. Therefore, we assume that the "new
geometrical quantum mechanics" (if exists) have to begin with the
finding out of geometrical interpretation of the basic quantum
equations whose validity is beyond question, and we start with the
attempts of finding out geometrical interpretation of the Dirac
equations. Some preliminary results were already published [6-10].

{\bf 1. Topological interpretation of the Dirac wave equation for
the free particle}

This equation (equation for the free particle with spin $1/2$) can
be written in the following symbolic form [11]
$$i\gamma^{\mu}\partial_{\mu}\psi=m\psi,\eqno (1)$$
where $\partial_{\mu}=\partial/\partial x_{\mu},\quad \mu
=1,2,3,4$,\quad $\psi (x)$ is the four-component Dirac bispinor,
$x_{1}=t, x_{2}=x, x_{3}=y$, $x_{4}=z$, and $\gamma^{\mu}$ are the
four-row matrices
$$
\gamma^0=\left(\begin{array}{cc}0&1\\1&0\end{array}\right), \quad
\gamma^{1,2,3}= \left(\begin{array}{cc}0&-{\bf s }\\{\bf
s}&0\end{array} \right),
$$ where ${\bf s}$ stands for the two-row Pauli matrices (we represent here four-row matrices trough two-row ones).
The summation in Eq.(1) goes over the repeating indices with a
signature $(1,-1,-1,-1)$. Here, $\hbar =c=1$. For the definite
values of 4-momentum $p_{\mu}$, the solution to Eq.(1) has the form
$$\psi=\psi_{p}\exp (-ip_{\mu}x^{\mu}).\eqno(2)$$ Substitution of (2) in Eq.(1) gives the following relation for  $p_{\mu}$
$$p_{1}^{2}-p_{2}^{2}-p_{3}^{2}-p_{4}^{2}=m^{2}.\eqno(3)$$

Within the modern interpretation of Eq.(1), solution (2) represents
a real physical object---classical wave field with mass and spin. As
a classical electromagnetic field, this field (as other possible
wave fields) may be represented, after quantization, as an ensemble
of quantum--elementary objects of the nature [11]. It is important
to emphasize that the space-time is considered within such
interpretation only as a scene where the above quanta exist and
interact. Below we will show that Eq.(1) may be interpreted as a
relation describing objects that are parts of the space itself. 

Being or not the solution to Eq.(1), four components of the Dirac
bispinors realize one of irreducible representations of the
space-time symmetry group--Lorentz extended group (4-rotations and
space inversion). Let us show that, being a solution of Eq.(1), the
bispinor (2) realizes the representation of one more symmetry group,
that was not discussed before and which has nothing to do with the
space-time symmetry. It is helpful to rewrite Eq.(1) in the form
$$
(l_{1}^{-1}\gamma^{1}T_{1}-l_{2}^{-1}\gamma^{2}T_{2}-l_{3}^{-1}\gamma^{3}T_{3}-l_{4}^{-1}\gamma^{4}T_{4})\psi
=l_{m}^{-1}\psi,\eqno(4)$$ where operators $T_{\mu} (\mu =1,2,3,4)$
have the form
$$T_{\mu}=i(2\pi)^{-1}l_{\mu}\partial_{\mu},\quad
l_{\mu}=2\pi p_{\mu}^{-1},\quad l_{m}=2\pi m^{-1}.\eqno(5)$$ We
rewrite solution (2) as
$$\psi=\psi_{p}\exp (-2\pi
ix_{1}l_{1}^{-1}+2\pi ix_{2}l_{2}^{-1}+2\pi ix_{3}l_{3}^{-1}+2\pi
ix_{4}l_{4}^{-1}).\eqno(6)$$ We also rewrite relation (3) as
$$l_{1}^{-2}-l_{2}^{-2}-l_{3}^{-2}-l_{4}^{-2}=l_{m}^{-2}.\eqno(7)$$
Note that all quantities in Eqs.(4--7) have the dimensionality of
length.

Operators $T_{\mu}$ in (5) and function $\psi (x)$ in (6) are
related by the following equation
$$\psi^{'}(x^{'}_{\mu})\equiv T_{\mu}\psi(x_{\mu}^{'})=\psi(x_{\mu}),\quad
x_{\mu}^{'}=x_{\mu}+l_{\mu}.\eqno(8)$$ This means that $T_{\mu}$ is
the operator representation of a group of parallel translations
along the $x_{\mu}$  axis over a distance $l_{\mu}$, and solution
(6) realizes this representation [12]. Being a four-component
spinor, $\psi(x)$ is related to the matrices $\gamma^{\mu}$ by the
equations (see, for example, [13])
$$\psi^{'}(x^{'})=\gamma^{\mu}\psi (x),\eqno(9)$$ where $x\equiv
(x_1,x_2,x_3,x_4)$, and $x^{'}\equiv (x_1,-x_2,-x_3,-x_4)$ for $\mu
=1,x^{'}\equiv (-x_1,x_2,-x_3,-x_4)$ for $\mu =2$, and so on. This
means that the matrices  $\gamma^{\mu}$ are the matrix
representation of the group of reflections along three axes
perpendicular to the $x_\mu$ axis, and the Dirac bispinors realize
this representation.

A parallel translation with simultaneous reflection in the
directions perpendicular to the translation is often spoken of as
"sliding symmetry" (see, for example, [14]). Thus, we see that the
operators
$$P^{\mu}=\gamma^{\mu}T_{\mu} \eqno(10)$$
form the representations of sliding symmetry group (sliding
symmetries in the $0x_{\mu}$ directions). Using the above notation,
we can rewrite Eq.(1) as a group-theoretical relation
$$l_{\mu}^{-1}P^{\mu}\psi=l_{m}^{-1}\psi,\eqno(11)$$ where, as we
showed before, solution (2) to this equation realizes the above
sliding symmetry group representation and, within suggested
interpretation, bears no relation to any wave processes into the
space--time.

The physical Minkovskii space-time has no sliding symmetry.
Therefore, we suppose that the above sliding symmetry is the
symmetry of some auxiliary 4-space and that the Dirac equation is
written in this space, whose symmetry describes symmetry of some
geometrical object. Such auxiliary spaces are used in topology for
the mathematical description of closed manifolds, because discrete
groups operating in these spaces (universal covering spaces of
manifolds) are isomorphic to fundamental groups of manifolds [15].
At the present time, only two-dimensional closed manifolds are
classified in details, and their fundamental groups and universal
covering planes are identified. Four-dimensional manifolds with the
above sliding symmetry group operating in pseudoeuclidean universal
covering space were not considered before. Therefore, we can only
use low-dimensional analogies, and two-dimensional nonorientable
closed manifold homeomorphic to the Klein bootle may be considered
as a possible such analogy, because its fundamental group is
generated by two sliding symmetries on the euclidean universal
covering space [14]. So, we assume that Eq.(1) describes a
four-dimensional nonorientable closed space-time manifold, and that
this manifold is the quantum object represented by the Dirac
equation. Nonorientable character of the manifold corresponds to the
fact that this object "has spin $1/2$" (note that spinors are just
the tensors that represent nonorientable geometrical objects).

First of all, let us show that the above assumption does not
contradict to our representation of physical objects as "something"
moving in space. Namely, we will show that assumed "geometrical"
quantum object looks like a propagating and increasing piece of
curved space. To get a better insight into the problem, we will use
below a one-dimensional analogy of our phenomenon. Let us consider
the simplest example of a closed topological
manifold---one-dimensional manifold homeomorphic to a circle whose
perimeter length is fixed and equal to $l_{0}$. This restriction on
the perimeter length plays here a role of restrictions (7). Such a
manifold is representable by all possible deformations of this
circle that conserves its continuity.

Let us show that the above manifold can be identified by the linear
differential equation that could be considered as an analogy to the
Dirac equation (1). The fundamental group of this manifold is a
group isomorphic to the group of integers [14]. This group is
isomorphic, in its turn, to the discrete group of one-dimensional
translations along a straight line over a distance $l_{0}$. This
line (0X axis) is the universal covering space of our manifold.
Therefore, the universal covering space for our circle is a
one-dimensional euclidean space where the above symmetry group
operates.

As we saw in (4--7), the operator
$$T_{x}=i(2\pi)^{-1}l_{0}d/dx \eqno (12)$$
is the operator representation of the above group of parallel
translations along 0X axis over the length $l_{0}$, and the function
$$\varphi(x)=\exp(-2\pi i x l_{0}^{-1})\eqno (13)$$
realizes this representation. This function satisfies the equation
$$i d\varphi/dx=m_{0}\varphi,\quad m_{0}=2\pi l_{0}^{-1}.\eqno (14)$$
Thus, linear differential equation (14) can be considered as one of
the possible descriptions of our one-dimensional manifold, and,
within the geometrical interpretation, this equation can be
considered as a very simplified analogy to the Dirac equation (1).

We will now show that our one-dimensional manifold describes
propagation of the pieces of the distorted one-dimensional space.
For simplicity, we consider here only all possible manifold
deformations that have a shape of ellipse with perimeter length
$l_{0}$. The equation for the ellipse on an euclidean plane has the
form
$$X^2/a^2+Y^2/b^2=1, \eqno (15)$$where all possible values of the
semiaxes $a$ and $b$ are connected with the perimeter length $l_{0}$
by the known approximate relation
$$l_{0}\simeq \pi[1,5(a+b)-(ab)^{1/2}].\eqno (16)$$This means that the
range of all possible values of $a$ is defined by the inequality
$0\leq a\leq a_{max}\simeq l_{0}/1,5\pi$.

In the pseudoeuclidean two-dimensional "space-time," the equation
for our ellipses has the form
$$X^2/a^2-T^2/b^2=1, \eqno (17)$$ and this equation defines the
dependence on time $T$ for the position of every point $X$ of the
manifold. At $T=0,X=\pm a$; that is, our manifold is represented by
the point set in one-dimensional euclidean space, and the dimension
of this point set is defined by all possible values of $a$. So, at
$T=0$, the manifold is represented by an extanded region of the
one--dimensional euclidean space $-a_{max}\leq X=a\leq a_{max}$. It
can easily be shown that at $T\neq 0$ this region increases and
moves along the X-axis. Thus, we can suppose that the Dirac equation
(1) describe propagation of an increasing region of the curved
euclidean space. The particular features of this propagation will be
considered elsewhere.

Finally, our hypothesis can be formulated as follows: {\it The Dirac
equation (1) is, in fact, the relation that describes the
topological and metrical properties of the microscopic closed
nonorientable space-time 4-manifold whose fundamental group is the
sliding symmetry group. Namely,it imposes limitations (7) on the
possible values of $l_{\mu}$---the lengths of paths belonging to
four different classes of the above fundamental group. The lengths
$l_{\mu}$ are defined trough 4-momentum by Eq.(5)  The notion of
"spin $1/2$" corresponds to the nonorientable character of the
manifold. Equation (1)is written not in the physical space-time but
in the auxiliary 4-space---universal covering space of the manifold,
where the above group operates, and this space coincides formally
with the Minkovskii space-time. "Wave" function (2) is a basic
vector of the fundamental group representation, and it bears no
relation to any wave field in the space. In the euclidean 3-space,
the above manifold appears as a propagation of the increasing piece
of a curved space}.

{\bf 2. Topological interpretation of the Dirac equation for a
hydrogen atom}

The suggested geometrical interpretation of Eq. (1) can be
considered as the "kinematic" hypothesis. To be approved, it should
be verified within the dynamic problems---quantum electrodynamics,
atomic spectra, and so on. In this Section we start with the
simplest dynamic problem where the interaction is the interaction
with a given static field. Namely, we will show that the Dirac
equation for a hydrogen atom allows topological interpretation as
the equation for free Dirac field.

The Dirac equation for hydrogen atom has the form [11]
$$i\gamma^{\mu}(\partial_{\mu}-ieA_{\mu})\psi=m\psi.\eqno (18)$$
Here $e$ and $m$ are charge and mass of an electron, $A_{\mu}$ are
electromagnetic potentials.

It was earlier shown by Fock that the expression in (18)
$$(\partial_{\mu}-ieA_{\mu})\psi $$
can be considered as a covariant derivative of the Dirac bispinors
in the special noneuclidean space (planar Weyl space) and that
electromagnetic potentials $ieA_{\mu}$ can be considered as a
connectivities of this space [2]. Up to now, the meaning of this
result was not clear, because physical space-time does not
demonstrate any features of the Weil space in the presence of
electromagnetic field. But Fock's result acquires a  physical
meaning only if we assume, on the basis of conclusions of previous
Section, that the equation (18) is written not in the physical
space, but in an auxiliary space---universal covering space of the
closed 4-manifold representing hydrogen atom.

Since the above result plays a key role, let us discuss properties
of the planar Weyl space in more detail. Geometry of this space is
specified by linear and quadratic forms [1]
$$ds^{2}=g_{ik}dx^{i}dx^{k}=\lambda (x)
(dx^{2}_{1}-dx^{2}_{2}-dx^{2}_{3}-dx^{2}_{4}),\eqno (19)$$
$$d\varphi=\varphi_{\mu}dx^{\mu},\eqno (20)$$
where $\lambda (x)$--is an arbitrary differentiable positive
function of coordinates $x_{\mu}$. This space is invariant with
respect to the scale (or gauge) transformations
$$g_{ik}^{'}=\lambda g_{ik},\quad \varphi_{i}^{'}=\varphi_{i}-\frac
{\partial \ln \lambda}{\partial x_{i}}.\eqno (21)$$ Therefore, a
single-valued, invariant sense has not $\varphi_{i}$ but the
quantity (scale curvature)
$$F_{ik}=\frac {\partial \varphi_{i}}{\partial x_{k}}-\frac {\partial
\varphi_{k}}{\partial x_{i}}.\eqno (22)$$ Antisymmetric tensor
$F_{ik}$ obeys equations that are analogous to the first pair of
Maxwell's equations
$$\partial_{i}F_{kl}+\partial_{k}F_{li}+\partial_{l}F_{ik}=0.$$

This analogy and the gauge invariance of  $\varphi_{i}$ (like the
gauge invariance of electromagnetic potentials) lead Weyl to the
idea that vectors $\varphi_{i}$ can be identified with the
electromagnetic potentials and that tensor $F_{ik}$ can be
identified with the tensor of electromagnetic field strengths
$$\varphi_{\mu}\equiv ieA_{\mu}, \quad A_{\mu}^{'}=
A_{\mu}-\partial_{\mu}\chi, \quad \chi = ie\ln \lambda.\eqno (23)$$
Then (like in general relativity), Weyl attempted to identify the
geometry of his space (curvature and so on) with the geometry of a
real space-time distorted by the presence of electromagnetic field
[1]. But it turned out that this hypothesis was contradictory to
some observable proprieties of the real physical space-time (it was
shown Einstein in the supplement to the Weyl publication [1]), and
the Weyl's results were afterwards considered as having nothing to
do with the electromagnetic field.

In contrast to Fock, we suppose that the covariant derivative in
(18) is written not in the real space-time but in the auxiliary
space--- universal covering space of topological manifold. So, there
are no objections against the Weyl space within our consideration.
This means that we can assume that the "long derivative" in (18) is
a covariant derivative written in the Weyl space and that the
4-potentials $ieA_{\mu}$ play the role of connectivities in the
above space. The concrete properties of the manifold representing a
hydrogen atom will be considered in subsequent publications, but
just now we can notice two important consequences of the topological
interpretation of Eq.(18).

1. It is known that connectivities of the Weyl space demonstrate the
same gage invariance as the gauge invariance of an electromagnetic
field [1]. This means that within the topological interpretation of
Eq.(18) the gage invariance of electromagnetic potentials $A_{\mu}$
is not some additional theoretical principle but is a natural
consequence of geometrical approach.

2. Geometrical interpretation of Eq.(18) for hydrogen atom does not
assume a presence of any point-like particles (electrons) inside the
atom. The wave function $\psi(x_{\mu})$ plays here the role of a
basic vector of the fundamental group representation. Coordinates
$x_{\mu}$ are coordinates of a point in the manifold universal
covering space, and this point bears no relation to some point-like
object. It seems reasonable to suppose that the same situation will
be realized within geometrical consideration of many-electrons
atoms. It is possible that the corresponding equations will turn out
to be the equations for functions of only one variable $x_{\mu}$,
and this will give the chance to overcome known difficulties of
many-body problem.

{\bf References}

\noindent1. Weyl H 1918 {\it Gravitation und Electrisitat, Berlin
Preus.Akad.Wiss}\\
2. Fock V 1929 {\it Zs. f. Phys.} {\bf 57} 261\\
3. Cheng T P, Li L F 1984 {\it Gauge theory of elementary particle
physics, Oxford, Clarendon Press, Ch.8}\\
4. Itzykson C, Zuber J B 1978 {\it Quantum Field Theory, New York
McGraw-Hill, Ch.12}\\
5. Wheeler J A 1960 {\it Neutinos, Gravitation and Geometry,
Bologna}
6. Olkhov O A, 2001 {\it Chem.Phys.Reports} {\bf 19} 1075\\
7. Olkhov O A, 2002 {\it Proc. Int. Conf. on the Structure and
Interactions of the Fhoton (Ascona Switzerland)} (New
Jersey-London-Singapure-Hong Kong: World Scientific) p 360\\
8. Olkhov O A 2002 {\it Topological interpretation of the Dirac
equation and geometrization of physical interactions} (Moscow:
Moscow Institute of Physics and Technology Press)\\
9. Olkhov O A, 2003 {\it Proc.XXIV Int.Colloquium on Group
Theoretical Methods in Physics (Paris)} (Group 24: Physical and
Mathematical Aspects of Symmetries Paris Institute of Physics
Conference Series Number 173, p 363) \\
10. Olkhov O A 2006 Geometrization of some quantum mechanics
formalism {\it Preprint} hep-th/0605060\\
11. Bjorken J D, Drell S D 1964 {\it Relativistic Quantum Mechanics
and Relativistic Quantum Fields, New York McGraw-Hill}\\
12. Hamermesh M 1964 {\it Group Theory and Its Application to
Physical Problems, Argonne National Laboratory}\\
13. Achiezer A I, Peletminski S V 1986 {\it Fields and Fundamental
Ineractions, Kiev Naukova Dumka, Ch.1}\\
14. Coxeter H S M 1961 {\it Introduction to Geometry, N.Y.-London
John Wiley and Sons}\\
15. Schvartz A S 1989 {\it Quantum Field Theory and Topology Moscow
Nauka}

\end{document}